\documentclass[twocolumn,showpacs,preprintnumbers,amsmath,amssymb]{revtex4}

\usepackage{epsfig}
\usepackage{graphicx}
\usepackage{dcolumn}
\usepackage{bm} 
\newcommand{\beq}{\begin{equation}}
\newcommand{\eeq}{\end{equation}}
\newcommand{\beqa}{\begin{eqnarray}}
\newcommand{\eeqa}{\end{eqnarray}}

\newcommand{\om}{\omega}

\def\jpb#1{{ J.\ Phys.\ B} {\bf#1}} 

\def\pra#1{{ Phys.\ Rev. A\/} {\bf#1}}

\def\prl#1{{ Phys.\ Rev.\ Lett.} {\bf#1}}

\begin{document}

\title{Effects of Elliptical Polarization on Strong-Field Short-Pulse
Double Ionization}

\author{Xu Wang}
\email{wangxu@pas.rochester.edu}
\author{J.\ H.\ Eberly}
\affiliation{ Rochester Theory Center and the Department of Physics 
\& Astronomy\\
University of Rochester, Rochester, New York 14627}

\date{\today}

\begin{abstract} We predict new end-of-pulse behavior in high-field atomic double ionization. Calculations of atomic electron trajectories in short intense laser pulses confirm our analysis of elliptical polarization. We exhibit a four-band structure in ion momentum distributions under various ellipticities, and predict that sequential and non-sequential double ionization can be cleanly distinguished under elliptical polarization. 
\end{abstract}


\maketitle


When atoms or molecules are exposed to an intense ($10^{14}$ to $10^{16}$ W/cm$^2$) femtosecond laser pulse, electrons that are released can collide under the laser force with the core. Many novel applications under the heading of attosecond physics are currently being vigorously pursued \cite{Krausz-Ivanov}. Among the effects of importance is double ionization. This was discovered experimentally to exhibit ion yields 4-6 orders of magnitude greater than predicted by the usual sequential tunneling theory. A characteristic ``knee" signature showing this enhancement has been observed with all rare gas atoms \cite{Fittinghoff-etal, Walker-etal, Augst-etal} and also some molecules \cite{Guo-Gibson, Cornaggia-Hering}. The  basis for the effect, i.e., for the exceptionally strong electron pair correlation, is the subject of active research, both experimentally and theoretically.  

\begin{figure}[t!]
\includegraphics[width=4cm,height=3cm]{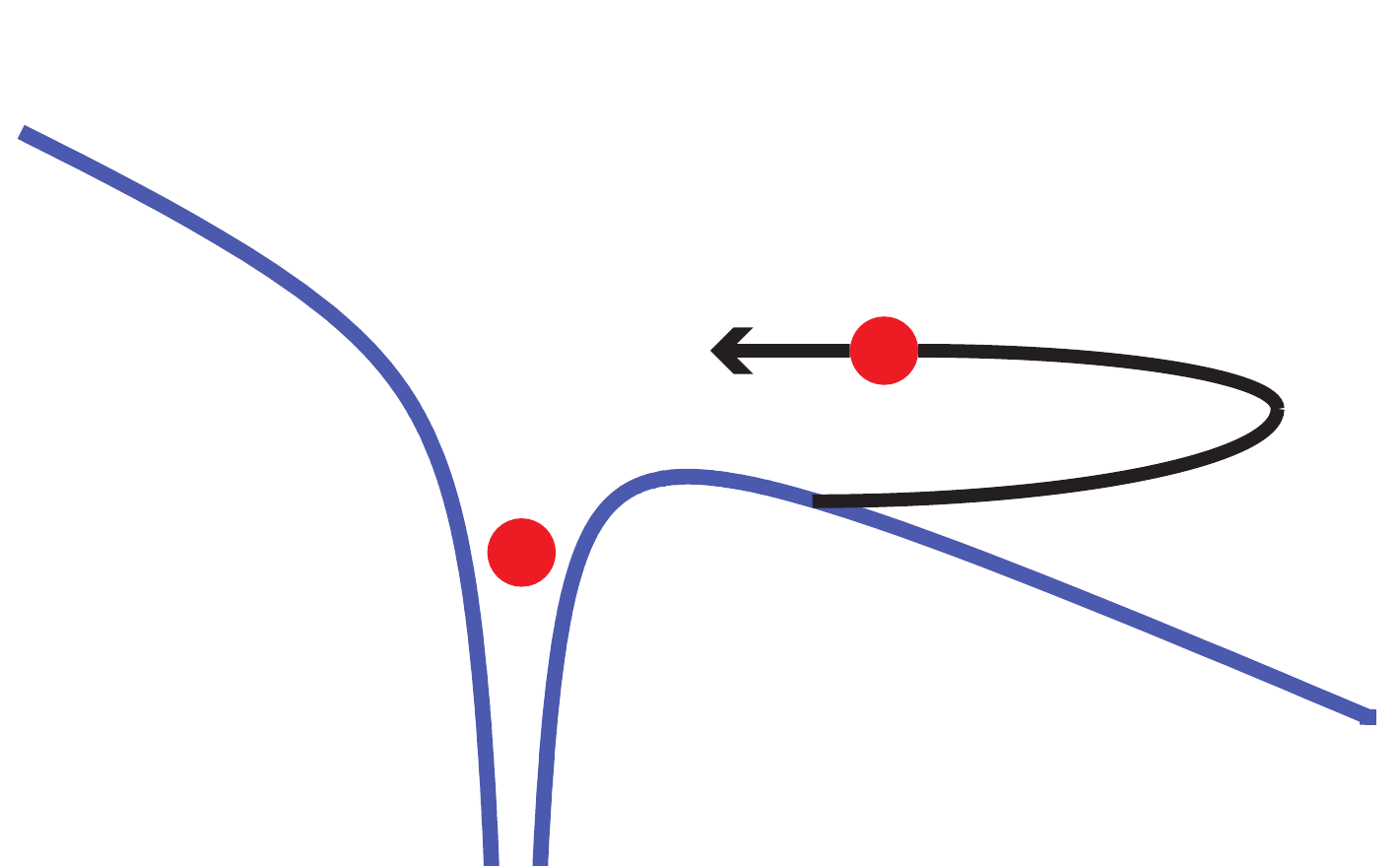}
\includegraphics[width=4cm,height=3cm]{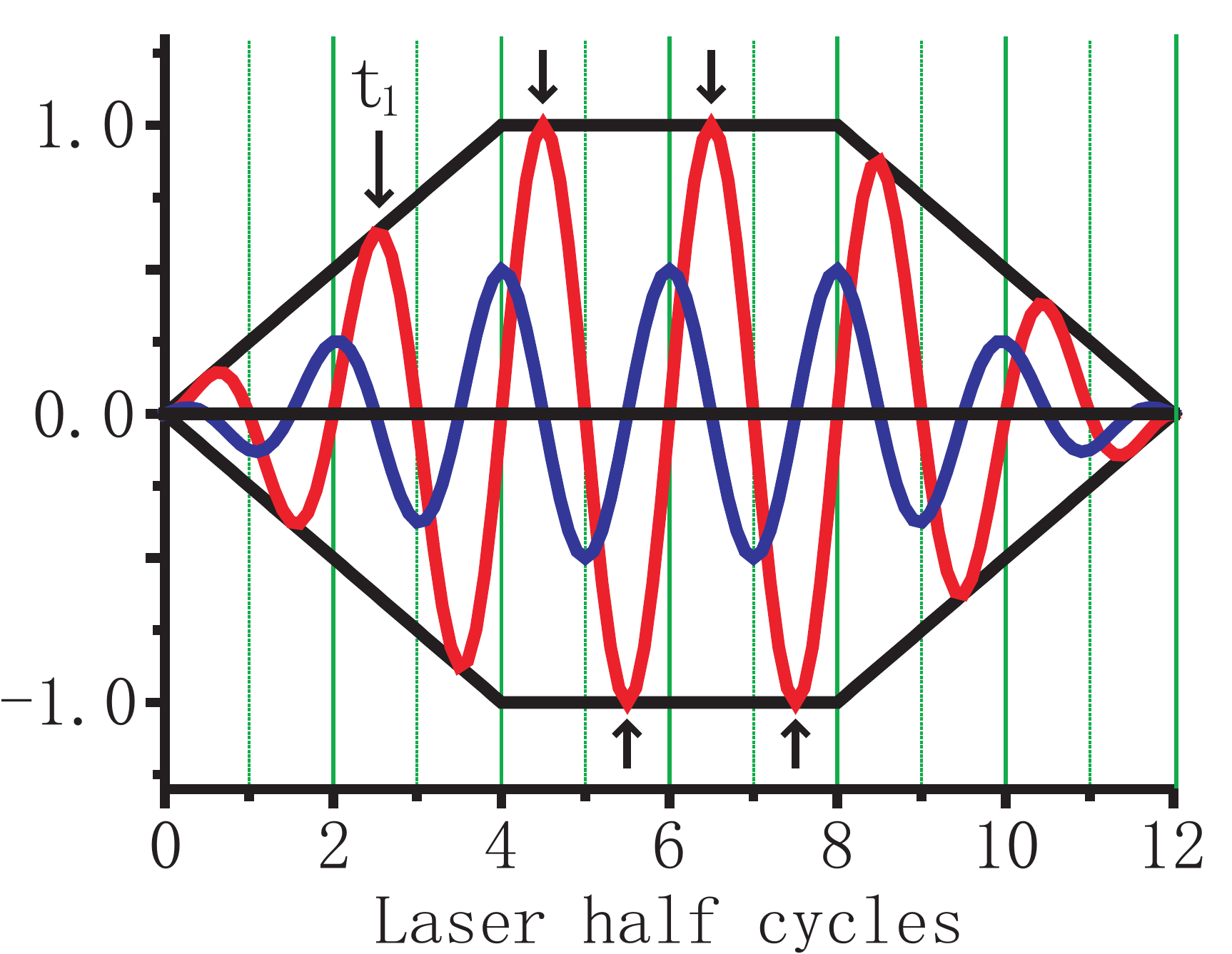}
\caption{{\footnotesize \label{f.pulse} The left panel shows the nature of the three-step recollision process for highly correlated double ionization. The right panel shows the flat-top model for the field given in Eqn. (\ref{e.Pol'dField}), here with $\phi=0$ and $\varepsilon$ = 0.5. The red and blue curves are the x and y components of the field. We take $\omega$ = 0.0584 a.u., i.e., $\lambda$ = 780 nm. The 10-cycle pulse used in our calculations would have a duration of about 26 fs. We choose $E_0$ = 0.131 a.u., corresponding to a peak (not average) intensity of $I$ = 0.6 PW/cm$^{2}$.  }}
\end{figure}

A three-step mechanism \cite{Corkum} illustrated in Fig. \ref{f.pulse} has generally been accepted as descriptive of such double ionization events: (i) one electron is freed with zero kinetic energy by tunneling when the nuclear coulomb potential has been tipped down by the strong laser field; (ii) the freed electron accelerates away in the laser field and is then driven back when the field reverses phase; and (iii) in returning, the laser-controlled electron acquires enough energy to carry away a second electron in a near-core e-e collision process. 

Many questions remain unsettled regarding strong-field double ionization, and one that is completely open concerns polarization. Almost all near-optical-frequency double ionization experiments have been carried out with linearly polarized light. The three-step picture suggests dramatically lower ion yield under elliptical or circular polarization, because the return trajectory is much less likely to encounter the atomic core, and this was quickly affirmed experimentally \cite{Dietrich-etal}. Double ionization has nevertheless been reported even with circular or near-circular polarization in studies of atomic magnesium \cite{Gillen-etal} and of several molecules \cite{Guo-Gibson}. 

The lack of cylindrical symmetry has up to now put the case of elliptically polarized fields beyond the range of essentially all quantum calculational approaches that are relevant to high-field double ionization \cite{Becker-Rottke, PPT}. However, with the recent development of two- and three-dimensional approaches to high-field ionization problems via classical ensembles \cite{ClassicalEnsemble}, a viable theoretical avenue is open \cite{Ho-Eberly06, Haan-etal06, Ho-Eberly07, Haan-etal08, Mauger-etal09}.  We report here surprising results from the first systematic theoretical exploration of double ionization with elliptically polarized light: 
\beq \label{e.Pol'dField}
\vec{E}(t) = E_0f(t)[\hat{e}_x\sin(\om t+\phi)+\hat{e}_y\varepsilon \cos(\om t + \phi)],
\eeq
as shown in Fig. \ref{f.pulse}.

Since multiphoton experiments in this intensity regime examine recoil ion momentum distributions \cite{Becker-Rottke, Weber-etal, Moshammer-etal, Feuerstein-etal, Eremina-etal, de Jesus-etal} as well as ion yield, we have calculated numerically the double ionization momentum distributions for a full range of ellipticities. A view of our classical ensemble results is provided in Fig. \ref{f.density} below. They significantly extend findings for non-zero ellipticity up to 0.5 reported by Shvetsov-Shilovski, et al. \cite{Shvetsov-Shilovski-etal}. We also present an analytical theory giving explanations for our numerical results and also for (a) an earlier speculative calculation \cite{Wang-Eberly09} and (b) previously unexplained features seen in very high field double ionization experiments on neon \cite{Cocke-etal}. We predict that the structures discovered should be observable with intensities and pulse durations readily available.

\begin{figure*}
\begin{center}
\includegraphics[width=4 cm, ]{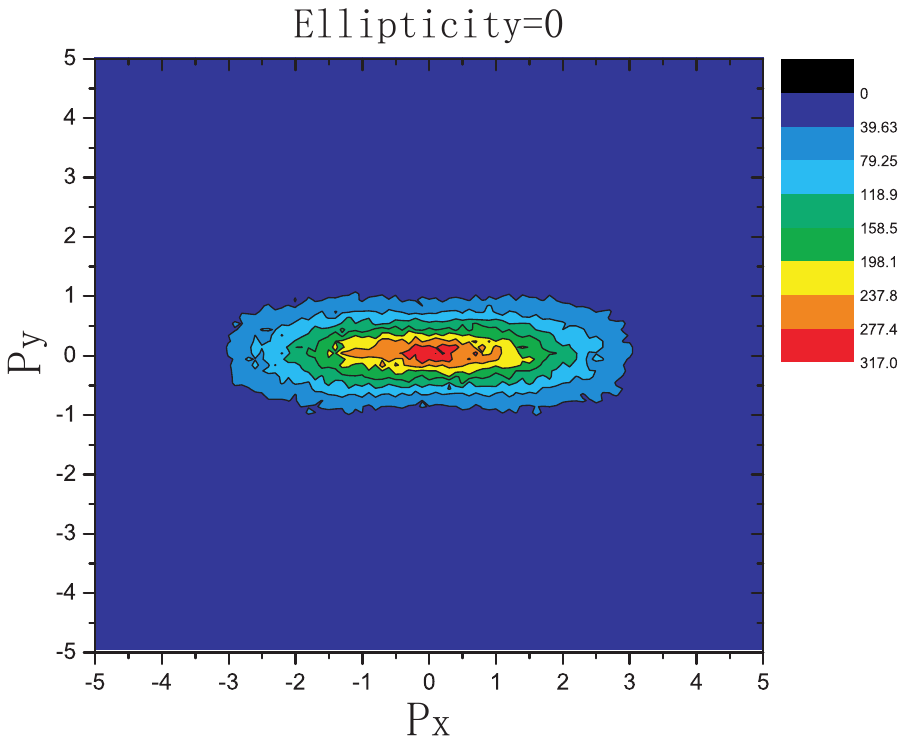}
\hspace{1cm}
\includegraphics[width=4 cm, ]{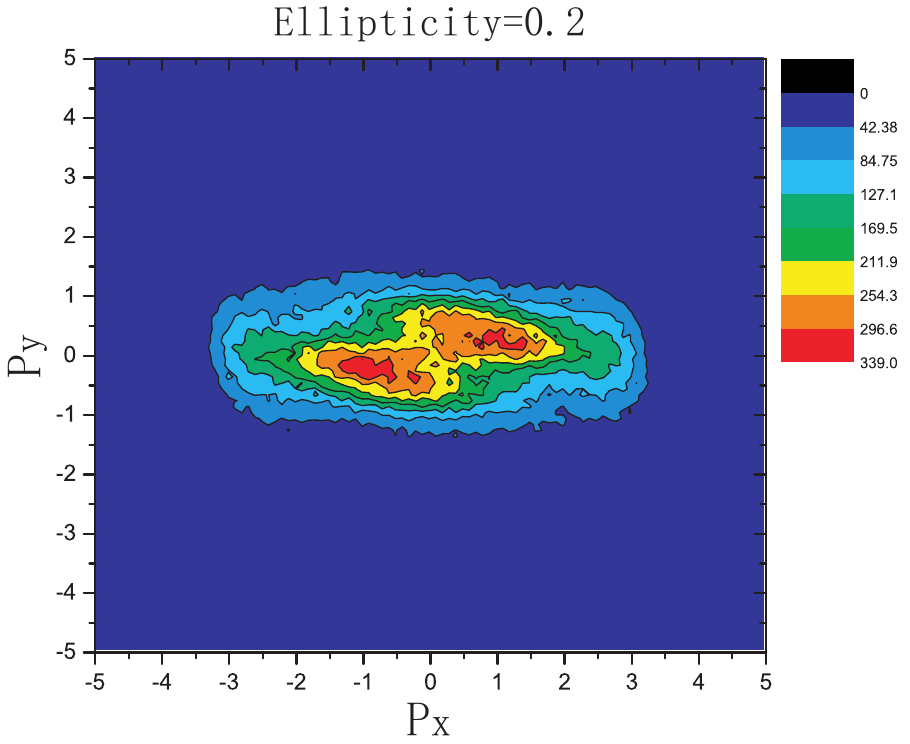}
\hspace{1cm}
\includegraphics[width=4 cm, ]{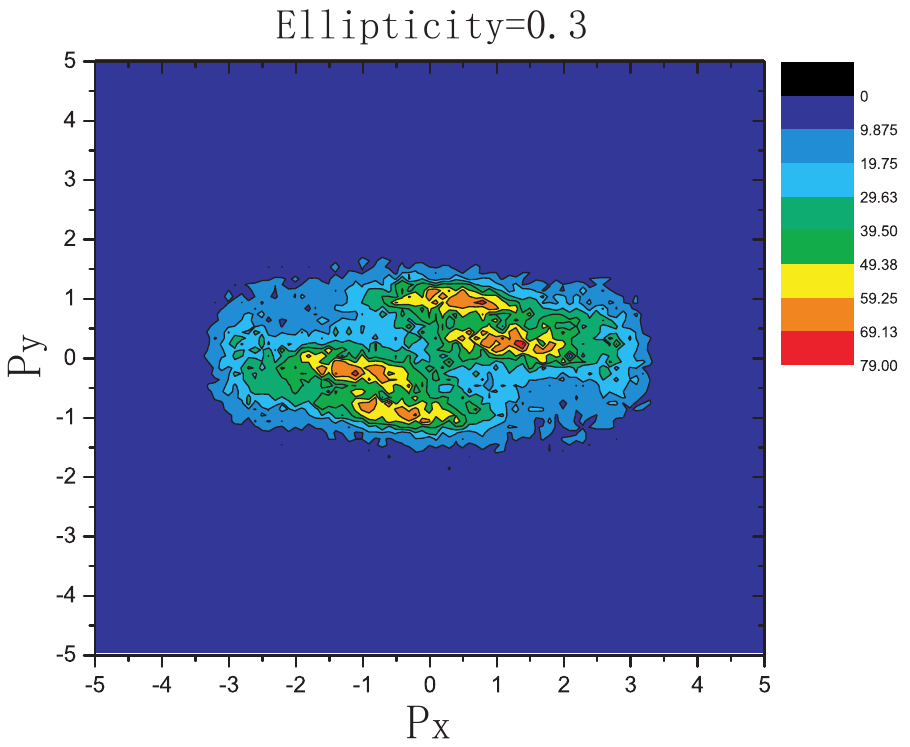} 
\includegraphics[width=4 cm, ]{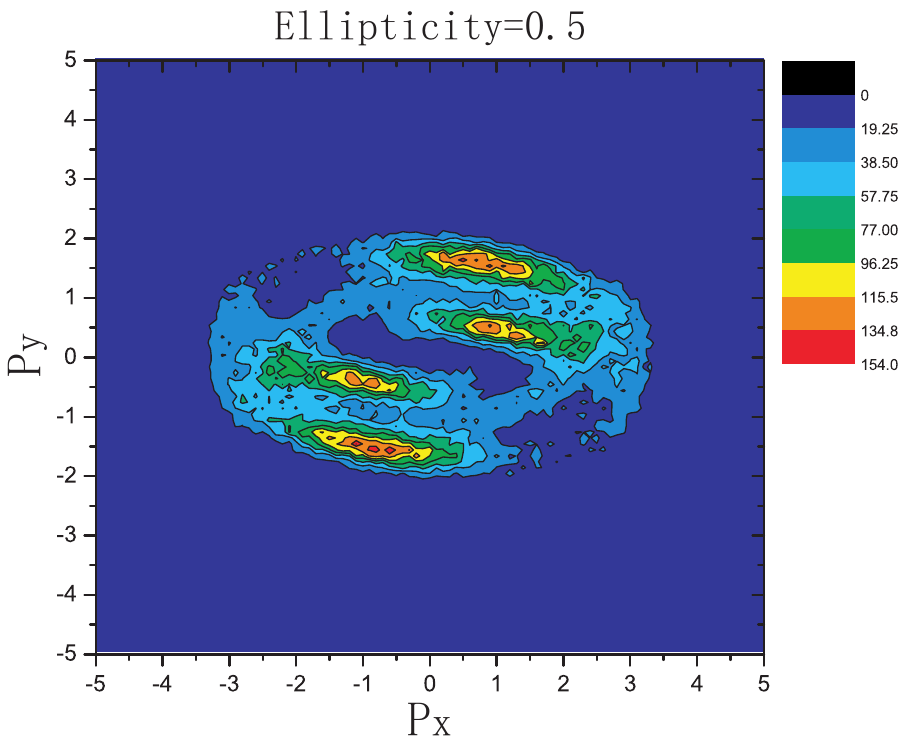}
\hspace{1cm}
\includegraphics[width=4 cm, ]{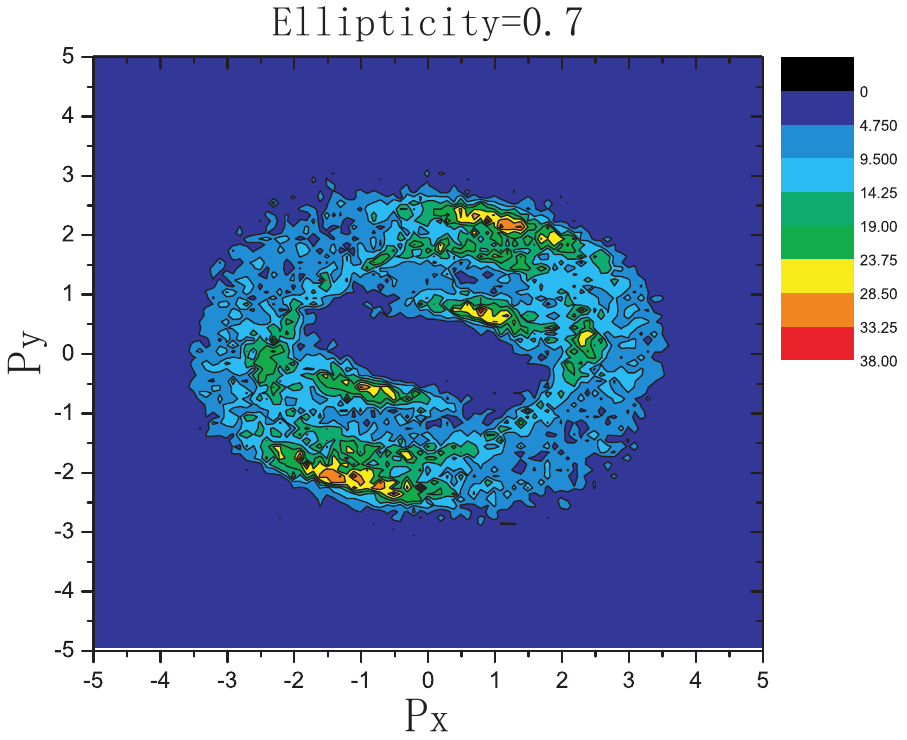}
\hspace{1cm}
\includegraphics[width=4 cm, ]{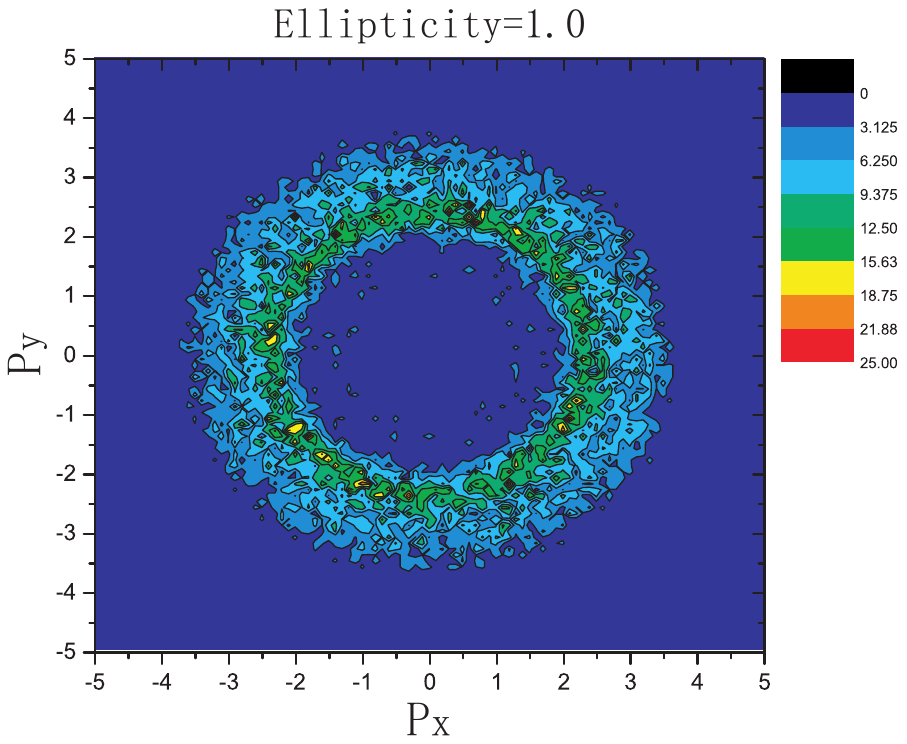}
\caption{{\footnotesize \label{f.density}  End-of-pulse momentum distributions for doubly ionized ions with ellipticities from 0 to 1, as indicated by the top label of each panel. }}
\end{center}
\end{figure*}

Our classical ensemble method has been described previously \cite{ClassicalEnsemble}, and it has already contributed to interpretation of double ionization phenomena \cite{TrajectoryTheory}. The $Z = +2$ charged ion core is fixed at the origin and in the energy of a two-electron bound system, $E_{tot} = \frac{1}{2}p_{1}^2 + \frac{1}{2}p_{2}^2 + V(\vec r_1,\vec r_2)$, we take 
$$V(\vec r_1,\vec r_2) = -ZV_{sc}(\vec r_1,a) -ZV_{sc}(\vec r_2,a) + V_{sc}(\vec r_1-\vec r_2, b),$$ 
with $V_{sc}$ standing for the soft-core Coulomb potential of the model \cite{Su-Eberly}:
\beq \label{e.scPotential}
V_{sc}(\vec r, c) = \frac{1}{\sqrt{r^2 + c^2}}.
\eeq
A many-pilot-atom method \cite{Abrines} is used to generate microcanonical ground state ensembles. In the absence of the laser field, the energy $E_{tot}$ of each 2e member of the ensemble is set to be -1.3 a.u. We expect that this value is close enough to the ground state energy of both Kr and Xe to be useful in guiding experiments  \cite{Energy-note}. The parameter $a$ can be regarded as the way the model accounts for the atomic core electrons. We set $a = 1.77$, to prevent autoionization, and $b = 0.1$, to allow strong electron pair correlation. Prior work with the model \cite{Ho-Eberly06} has shown that out-of-plane effects are negligible in first approximation at the intensities used experimentally, so we restrict the electrons to motion in the x-y polarization plane for simplicity. The end-of-pulse momentum distributions of doubly ionized ions, which are calculated as the sum of the momenta  of the two ionized electrons, are shown in Fig. \ref{f.density}. 

Let us look at each graph in detail. For linear polarization there is  one continuous region in the momentum domain, which lies along the $p_x$-axis. As ellipticity increases to 0.2, this region grows into two parts: the right-upper part that lies mainly in the first quadrant and the left-bottom part that lies mainly in the third quadrant. As ellipticity increases to 0.3, each of the two regions further grows into two new parts and a four-band structure emerges. At ellipticity 0.5, the four-band structure can be clearly seen. Besides, a new elliptical structure becomes obvious, part of which appears as two ``bridges" connecting the four-band structure. This elliptical structure grows and in the meantime the four-band structure decreases as ellipticity increases. At $\varepsilon =1$ the distribution is circularly symmetric.

\begin{figure}[b!]
\includegraphics[width = 3.5cm,height=3cm]{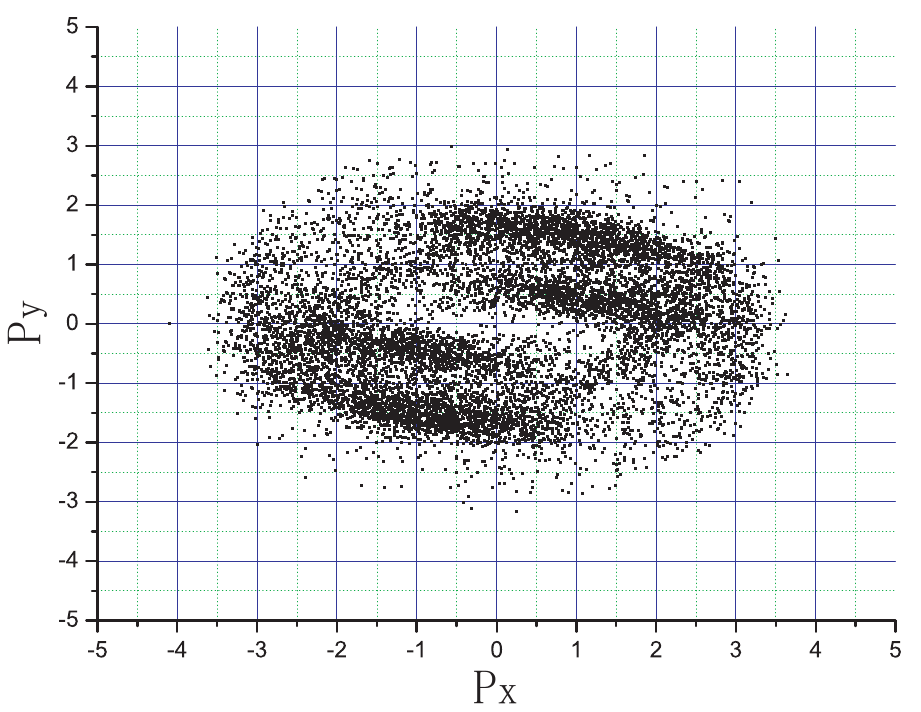} 
\hspace{2cm}
\includegraphics[width = 3.5 cm,height=3cm]{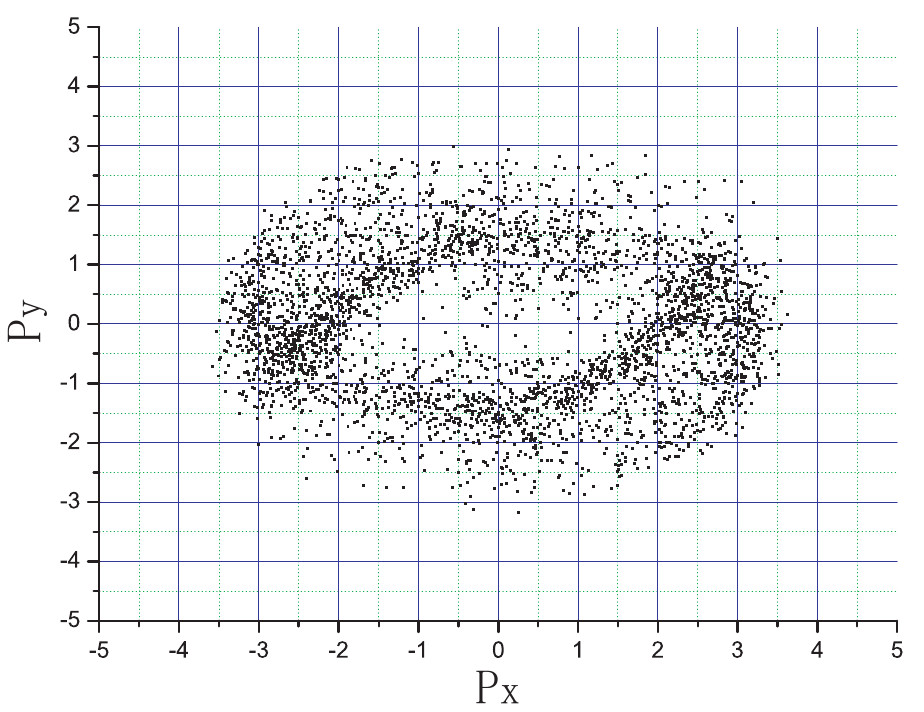}
\includegraphics[width = 3.5 cm,height=3cm]{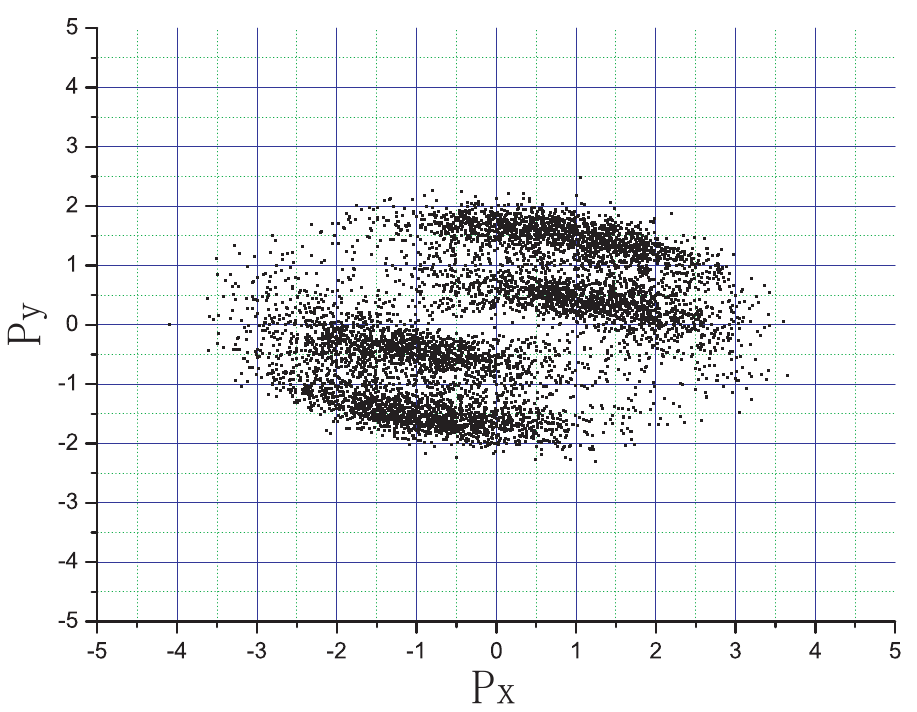} 
\caption{{\footnotesize \label{f.DI}  Momentum distribution of DI (top) can be further divided into momentum distribution of NSDI (left) and momentum distribution of SDI (right).  Ellipticity used here is 0.5.}}
\end{figure}

To determine the origin of these momentum distributions, the history of each two-electron trajectory was traced back. Trajectory back analysis \cite{BackAnalysis} provides great insight into physical processes and is only possible with a classical approach. Depending on whether there are close recollisions or not, events are classified as non-sequential double ionization (NSDI) or sequential double ionization (SDI), as shown in Fig. \ref{f.DI}. This figure is the first to show distinct patterns of ionization in which SDI and NSDI are separated cleanly by ellipticity. Our observation that in linear polarization 10\% of the single ionizations are converted to SDI events appears consistent with 1-D reports from the phase space perspective \cite{Mauger-etal09}. 

\begin{figure}[t!]
\includegraphics[width=3cm]{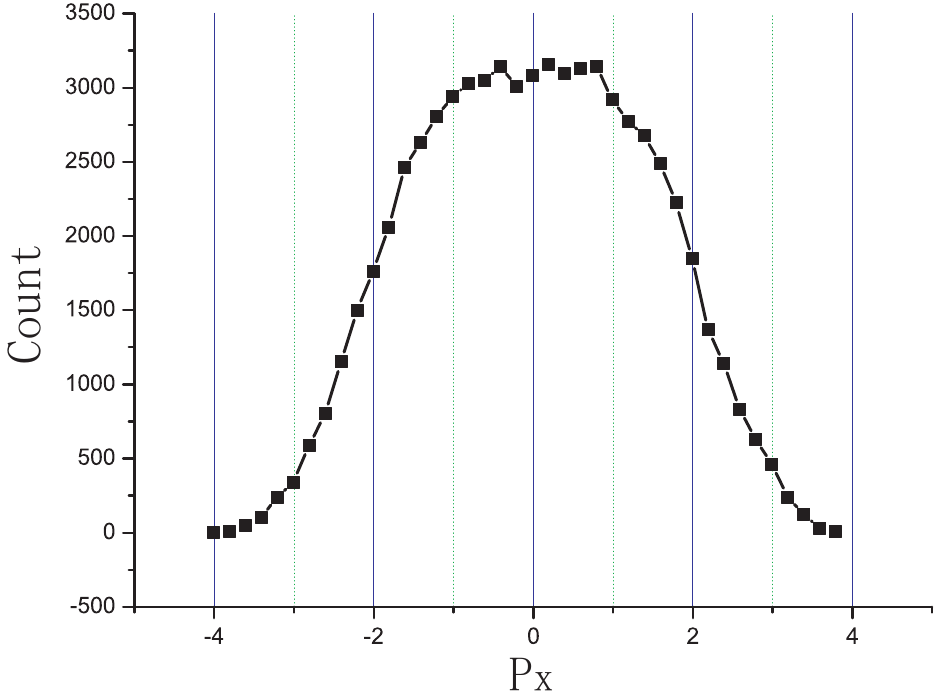} \hspace{0.5cm}
\includegraphics[width=3cm]{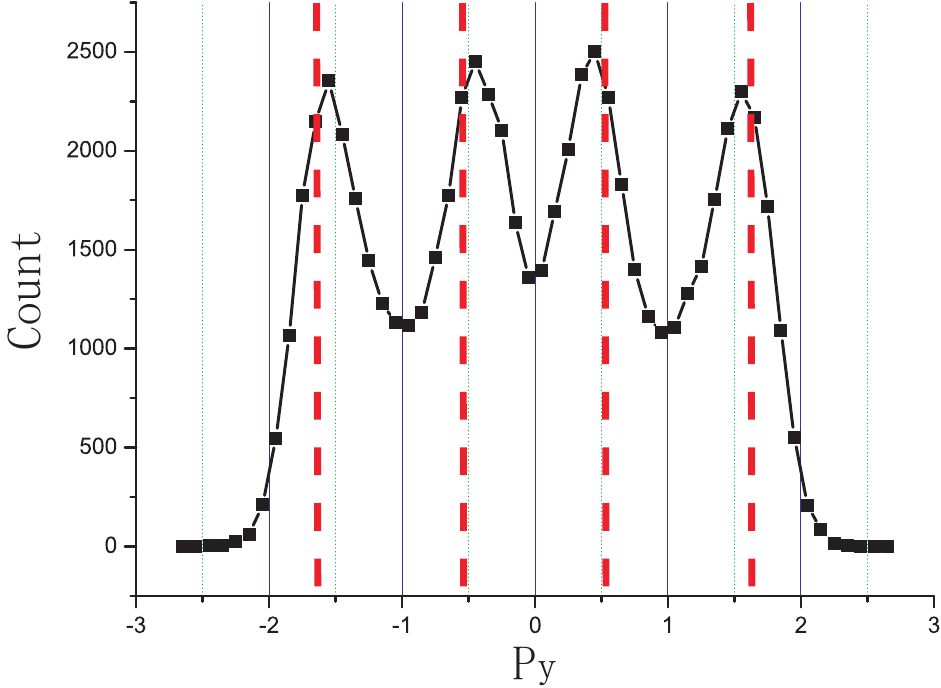}
\caption{{\footnotesize \label{f.SDI_xy} Momentum distribution of SDI in x-direction (left) and in y-direction (right). The $P_x$ distribution shows a single peak centered at zero while the $P_y$ distribution shows four-peaks. Four vertical dashed lines are positions of peaks predicted by our analysis (see Eqn. (\ref{e.DeltaPy})), which clearly fits well with the numerical results.}}
\end{figure}

One can see in Fig. \ref{f.DI} that SDI is responsible for the four-band structure and NSDI for the elliptical structure. It is helpful to project the 2D momentum distribution of SDI onto the x- and y-directions. The SDI distribution along the x-direction is a single peak structure centered at zero while the distribution along the y-direction has four peaks centered at approximately $\pm 0.5$ a.u. and $\pm 1.5$ a.u., as shown in Fig. \ref{f.SDI_xy}. This four-peak character has probably already been recorded without notice at least once in very high field sequential ionization of neon under nearly circular polarization (see the study by Maharjan, et al. \cite{Cocke-etal}).

\begin{figure}[b!]
\includegraphics[width=3cm]{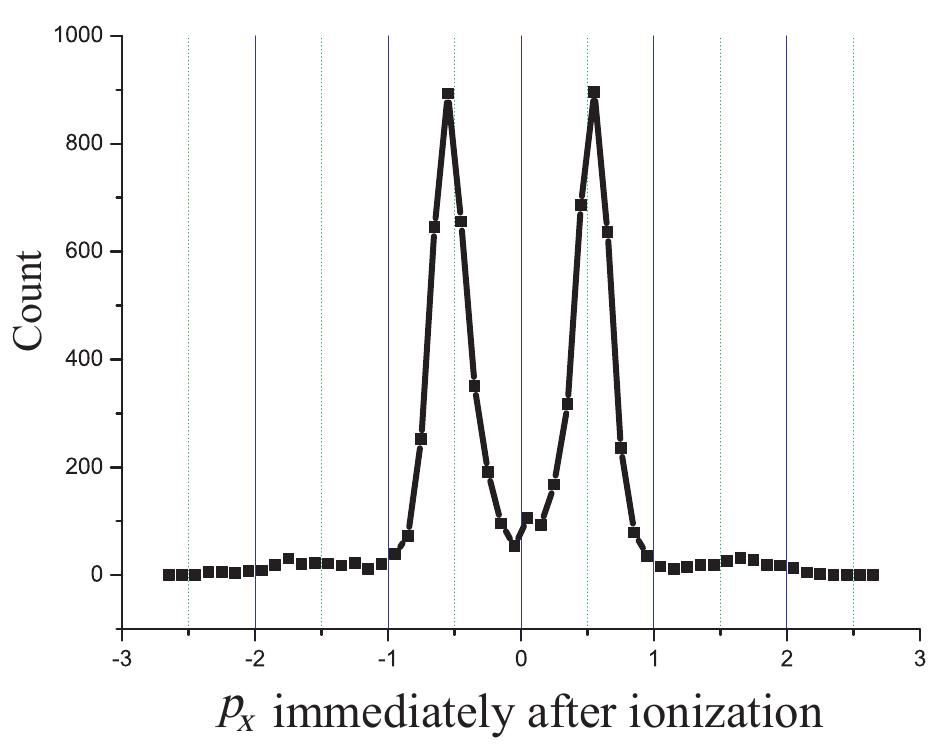}
\includegraphics[width=3cm]{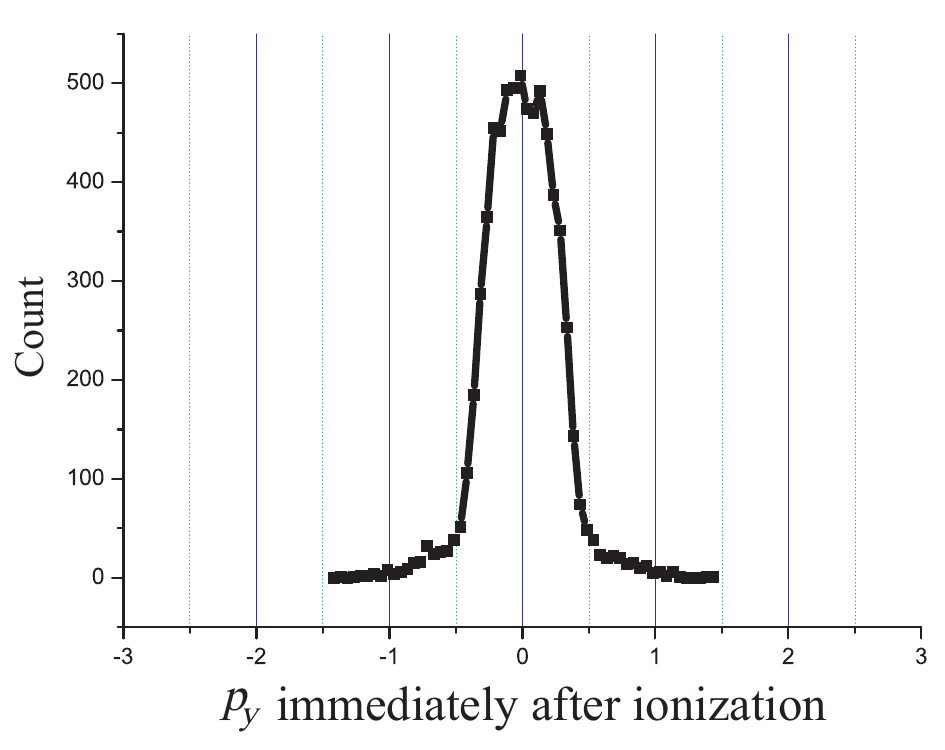}
\caption{{\footnotesize \label{f.MD_i} Momentum distributions of electrons (first or second) immediately after ionization along the x- and y- directions. It is a double peak structure centered at $\pm$ 0.5 a.u. along the x-direction while a single peak structure is centered at zero along the y-direction.}}
\end{figure}

Before beginning an analysis of the predicted four-peak structure one important feature needs to be pointed out. For single or sequential double ionization, electrons are almost always ionized \cite{Ionization-note} along the x-direction (plus or minus) when the field is near maximum because the peak field in the x-direction is higher than that in the y-direction, due to ellipticity. At those x-maximum times, the field in the y-direction is zero. As a consequence, the electron momentum distribution shows a double peak structure centered at $\pm$ 0.5 a.u. along the x-direction while a single peak is centered at zero along the y-direction, as indicated by Fig. \ref{f.MD_i}. 

After the electrons are ionized, the reasonable approximation can be made that the Coulomb attraction between the electron and the ion core can be neglected and the only force that acts on the freed electron is the laser force. Then the change of momenta of the two electrons due to drift motion with the laser field can be derived analytically. We let $t_{1}$ and $t_{2}$ be the times when the first and second electrons are ionized. Then the total change of momentum of the two electrons due to drift motion along the y-direction is
\beqa \label{e.DeltaPy}
\Delta P_{y} &=& \Delta p_{1y} + \Delta p_{2y} \nonumber\\
&=& \frac{E_{0} \varepsilon}{4 \pi \omega} [\omega t_{1} \sin(\omega t_{1}+\phi)+4\pi \sin (\omega t_{2}+\phi) -\cos \phi] \nonumber\\
&=& \frac{E_{0} \varepsilon}{4 \pi \omega} [\omega t_{1} \sin(\omega t_{1}+\phi) \pm 4\pi \sin (\omega t_{1}+\phi)-\cos \phi]. \nonumber
\eeqa
As we explain below, $\sin(\omega t_{1} + \phi) \simeq 1$ and the final $\cos\phi$ should be dropped, in which case we find four preferred values, to be compared with Fig. \ref{f.SDI_xy}:
\beq \label{e.DeltaPy-2}
\Delta P_{y} \approx 0.089 \times (2 \pi \pm 4\pi) \sin (\omega t_{1} + \phi).
\eeq

Now let us explain the origin of the terms in our analysis in detail, and evaluate them numerically. Because we are in a regime where a significant amount of SDI can be achieved, it is obvious that the first electron will be ionized before the plateau stage of the field, i.e., during the turning-on stage, since the first electron is much easier to ionize than the second one. This point is confirmed by our trajectory calculations. It turns out that most of the first electrons are ionized around the beginning of the second cycle, so $\omega t_{1}$ is in the near neighborhood of $2\pi$ and $\cos\phi$ averages to zero. The difference between $t_{2}$ and $t_{1}$ could be any number of half cycles -- $2\pi/\omega$, $3\pi/\omega$, $4\pi/\omega$ and so on, as indicated by arrows in Fig. \ref{f.pulse}. In the odd half-cycle cases, when $t_{2} - t_{1} = (2n+1)\pi/\omega$, the two electrons leave the ion core along opposite directions (along $+x$ and $-x$). The result has been called \cite{Ho-Eberly03} the ``Z" scenario, for approximately zero momentum transfer, or {\em out of phase} double ionization. In the even half-cycle cases, $t_{2}-t_{1} = 2n\pi/\omega$, the two electrons leave the ion core along the same direction (either $+x$ or $-x$) and we label this ``NZ" for substantially non-zero momentum transfer, or {\em in phase} double ionization.

\begin{figure}[t!]
\includegraphics[width=3cm]{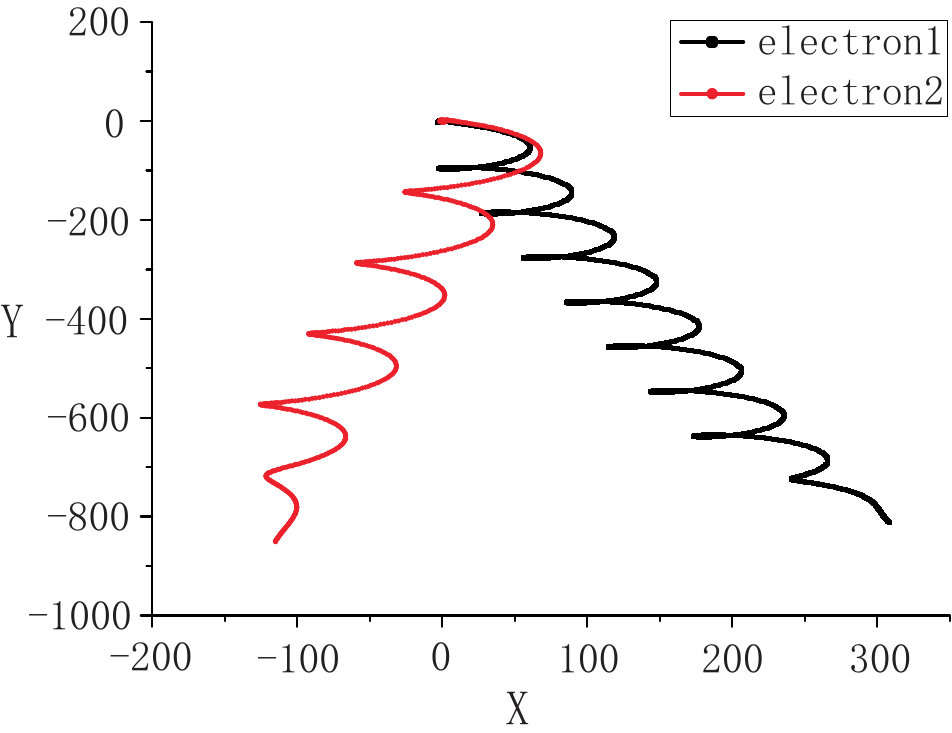}
\includegraphics[width=3cm]{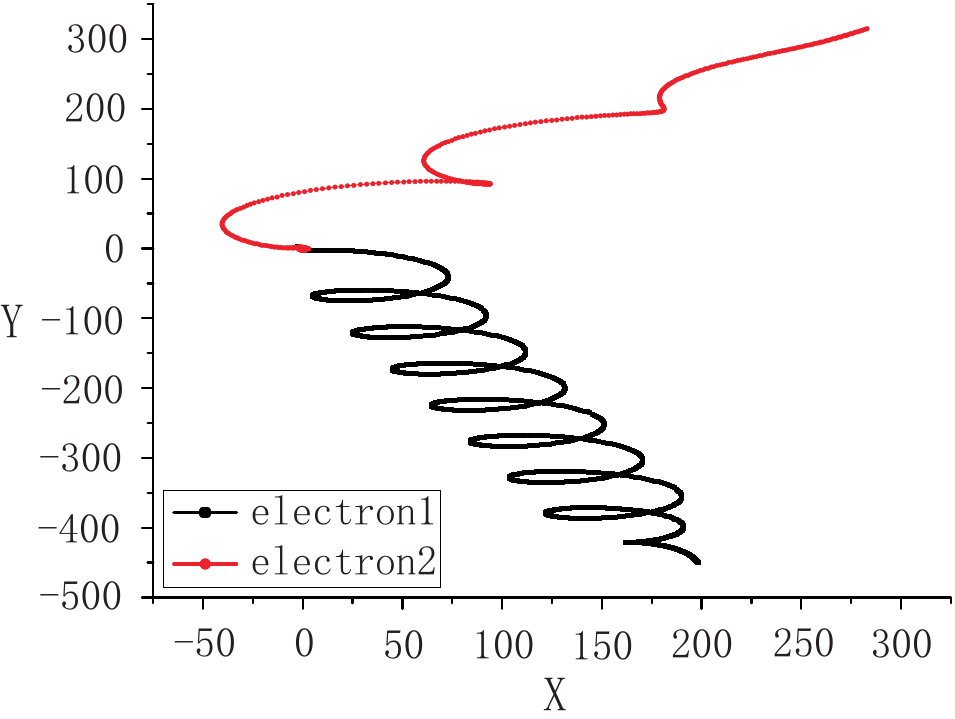}
\caption{{\footnotesize \label{f.trajectories} In-phase and out-of-phase (NZ and Z) trajectories are shown for clarity.}}
\end{figure}

The interesting y-components of the end-of-pulse momenta of the two electrons are explained as follows. For {\em in phase} SDI, the y-components of the end-of-pulse momenta of the two electrons most probably also have the same sign, as shown in Fig. \ref{f.trajectories}. Hence the magnitude of the y-component of the end-of-pulse momentum of the ion is large, which corresponds to the outer two peaks of the four-peak structure in Fig. \ref{f.SDI_xy}. For {\em out of phase} SDI  the y-components of the end-of-pulse momenta of the two electrons most probably also have opposite sign, hence the magnitude of the y-component of the end-of-pulse momentum of the ion is small, which corresponds to the inner two peaks of the four-peak structure. On the assumption that the electrons are ionized exactly when the field in the x-direction is maximum, we have $\sin(\omega t_{1} + \phi) = \pm 1$ and $\sin(\omega t_{2} + \phi) = \sin(\omega t_{1} + 2n\pi + \phi) = \sin(\omega t_{1}+\phi)$  (\emph{in phase}) or $\sin(\omega t_{2}+\phi)= \sin(\omega t_{1} + (2n+1)\pi + \phi) = -\sin (\omega t_{1}+\phi)$ (\emph{out of phase}). The peak positions predicted in Eq.(\ref{e.DeltaPy}) are $\pm 0.56, \pm 1.68$, close to matching the ensemble average values shown in Fig. \ref{f.SDI_xy}.

In summary, we have predicted a new end-of-pulse momentum distribution for sequential double ionization, and have shown that elliptical polarization provides a new control mechanism for recollision physics in  high-field ionization. Our trajectory calculations are made in the spirit of the standard recollision picture and are analytically feasible because of our special trapezoidal pulse shape. The numerical distributions are only possible to obtain classically at this time. This is because no workable non-perturbative quantum approach to elliptical high-field short-pulse effects is known. The good agreement achieved provides additional support for the interpretation of high-field ionization processes as examples of electron physics first, and atomic physics secondarily, a view that encourages classical interpretations and calculations of even more complex high-field effects. Finally we remark that the same four bands with the same separation are obtained numerically from short smooth Gaussian pulses as well.

We appreciate the contributions from Phay J. Ho at the early stage of this work, as well as communications with C. Guo, S.L. Haan and D.D. Meyerhofer. This work was supported by DOE Grant DE-FG02-05ER15713.


\begin{thebibliography}{99}
\bibitem{Energy-note}    The energy levels of Kr and Xe are calculated by summing the first and second ionization potentials. For Kr, the 1st and 2nd ionization potentials are 14.00eV and 24.36eV, which yields an energy level -(14.00+24.36)eV= -38.36eV= -1.41a.u.; For Xe, the figures are 12.13eV and 21.21eV, and an energy level -1.23a.u.
\bibitem{Ionization-note}  There is no standard criterion for ionization of an electron trajectory. One may define it to occur when the distance from the ion center is some chosen large value, e.g., 6 a.u. in our case. We have checked that small variations in definition of ionization do not lead to significant differences in the final result and do not affect our discussion here.
\end{thebibliography}
\end{document}